# Detecting and Explaining Changes in Various Assets' Relationships in Financial Markets

Makoto Naraoka[1], Teruaki Hayashi[1[0000-0002-1806-5852]], Takaaki Yoshino[2], Toshiaki Sugie[2], Kota Takano[2], and Yukio Ohsawa[1[0000-0003-2943-2547]]

[1] The University of Tokyo, 113-8656, 7-3-1, Hongo, Bunkyo-ku, Tokyo, Japan
[2] Nissay Asset Management Corp., Japan

**Abstract.** We study the method for detecting relationship changes in financial markets and providing human-interpretable network visualization to support the decision-making of fund managers dealing with multi-assets. First, we construct co-occurrence networks with each asset as a node and a pair with a strong relationship in price change as an edge at each time step. Second, we calculate Graph-Based Entropy to represent the variety of price changes based on the network. Third, we apply the Differential Network to finance, which is traditionally used in the field of bioinformatics. By the method described above, we can visualize when and what kind of changes are occurring in the financial market, and which assets play a central role in changes in financial markets. Experiments with multi-asset time-series data showed results that were well fit with actual events while maintaining high interpretability. It is suggested that this approach is useful for fund managers to use as a new option for decision making.

## 1 Introduction

This study aims to detect changes in the relationship between various assets in the global financial markets in a highly interpretive manner to support fund managers' investment decision-making. To achieve this objective, we construct and visualize a network based on the strength of the relationship between each asset, which changes dynamically at each point in time. Also, we calculate quantitative indices based on the network structure, which enables us to capture the overall trend in the financial market. The method combines visualization and quantification in a way that allows fund managers to examine what kind of changes are happening in the market and what factors are behind those changes, as well as to estimate the scale of those changes. We think that these will help fund managers to make investment decisions.

Anomaly detection of time series data has been actively studied [1-2] and is used in various fields. However, in most cases, the research is aimed at improving the accuracy of the detection, and not many studies have approached the clarification of the causes of the anomaly at the same time. When such algorithms are used in industry for decision making, it is often necessary to interpret why such changes and anomalies have occurred. Therefore, we consider that a methodology that combines change detection with visualization to understand the causes of the change will be particularly useful as a tool for decision making in industry, including finance.

Asset management companies and banks sell mutual fund financial products related to stocks and bonds. In Japan, for example, Japanese stocks, developed country stocks, Japanese bonds, and U.S. bonds are often sold according to the type of asset: stocks for stocks, bonds for bonds, and so on. In recent years, there has been a growing interest among investors in multi-asset funds, which combine multiple types of assets and allocate them according to market trends as needed. In response to this demand, asset management companies are also building and selling multi-asset funds. Multi-asset funds are often evaluated based on indicators that are calculated by correlations and diversification between assets [3-5], but there are not yet many indicators from other approaches; the dangers of looking at a few indicators have also been pointed out [6].

In this study, we propose indicators using Graph-Based Entropy with co-occurrence networks between assets and the number of hubs in differential networks. These indicators are based on a different approach than conventional correlation and variance-based indicators and are expected to enable the monitoring of multi-asset markets from a different perspective than conventional indicators. It also helps fund managers monitoring multi-asset funds to capture the changes occurring in the market by visualizing the market conditions in a network format through co-occurrence and differential networks. Graph-based anomaly detection and explanation research has been conducted in a variety of domains, and graph-based approaches are very effective in ensuring explainability and interpretability [7]. Although several studies analyze financial markets from a network approach [8-10], we do not find any studies that apply it to multi-asset markets. Graph-Based Entropy based on co-occurrence networks used in our method has been used in the field of marketing in the retail industry [11-12], and differential networks have been used in the field of bioinformatics [13-15]. This study is the first to apply them to finance. The results of applying this method to multi-asset consisting of the stock, bond, and foreign exchange markets in 2007 are consistent with real-world market changes, suggesting the effectiveness of this method.

## 2 Method

### 2.1 Co-occurrence Networks and Graph-Based Entropy

First, to calculate the distance matrices, we preprocessed the raw time-series data according to the following procedure. (ⅰ) divided the time series by window, (ⅱ) standardized (Z-score) within the window, and (ⅲ) corrected for price movement directionality based on fund managers' experience.

The standardization was performed as shown in Equation 1, where t is the time and w is the window width.

$$Zscore_t = \frac{x_t - average\left(x_{t-w,t-w+1,\dots,t}\right)}{SD\left(x_{t-w,t-w+1,\dots,t}\right)} \quad (1)$$

The directional corrections are made by multiplying those that tend to increase in price at risk-on by 1 and those that tend to decrease by -1. For example, stocks have high volatility and are a high-risk, high-return asset, so investors tend to buy them aggressively when the market is strong for high profits. Bonds, on the other hand, have low volatility and are unlikely to lose their principal unless the country defaults, but they are less profitable than stocks. Conversely, when the outlook for financial markets is uncertain, investors tend to abandon stocks and buy bonds. Based on this rule of thumb stock price is multiplied by 1 because the price tends to rise when the risk is on, and bond price and exchange rates are multiplied by -1 because they tend to fall in value.

Based on the preprocessed time-series data for each asset, we calculate a distance matrix that represents the degree to which the price movements of each asset are similar. We used Dynamic Time Warping (DTW) [16] to calculate the distance from an asset to another. DTW is a method of stretching the time axis so that the distance between two sequences is minimized. It is often used as a distance calculation method for time-series data because it is more noise-resistant than Euclidean distance and more suitable for human intuition. The distance between the sequence $P = \{p_1, \cdots, p_l\}$ and the sequence $Q = \{q_1, \cdots, q_m\}$ is below in Equation 2, and the distance can be calculated by matching each element of the sequence P with each element of the sequence Q in ascending order.

$$D_{dtw}(P,Q) = f(p_l, q_m),$$
$$f(i,j) = \left\|p_i - q_j\right\| + \min \begin{cases} f(i,j-1) \\ f(i-1,j) \\ f(i-1,j-1) \end{cases} \quad (2)$$

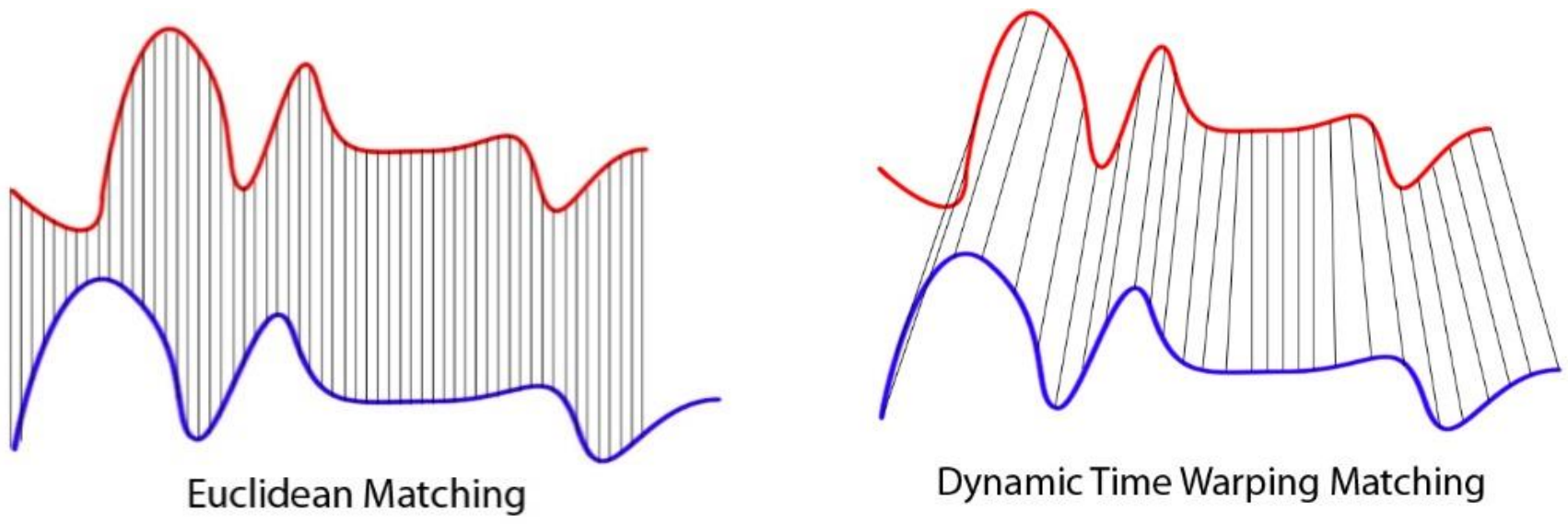

(**a**) (**b**)

**Fig. 1.** (a) The distance measured by Euclidean (b) The distance measured by DTW. [partially modified by authors], Source: XantaCross, (2011) "Euclidean vs DTW", Retrieved May 21, 2020, from https://upload.wikimedia.org/wikipedia/commons/6/69/Euclidean_vs_DTW.jpg

In this study, the time series data pairs used to calculate distance are multi-asset, and time series data with different measurement timings are sometimes compared, and time scaling methods such as DTW have an advantage in this case. For example, the assets in this study include those from different regions, such as Japanese and U.S. stocks. In the case of Japan and the U.S., price movements may be mutually influential due to the strong economic ties between the two countries. Besides, prices can be affected by a lag due to differences in trading hours caused by time differences. Of course, Japanese stocks may also move in price on their own due to factors unique to Japan. We expect that DTW will be able to handle such time differences automatically and appropriately.

Based on the distance matrixes, if the distances for each pair of assets are smaller than a certain threshold, we consider them to be similar and create and visualize the network by putting an edge on them. By the above, a co-occurrence network can be constructed.

After constructing the Co-occurrence Network, we quantify the network with Graph-Based Entropy (GBE) [11]. GBE is a measure of Shannon's entropy applied to the network structure, calculated based on the clusters represented in the network and defined as follows.

$$\mathrm{Hg} = -\sum_j p(cluster_j) \log_2 p(cluster_j)$$
$$where\ p(cluster_j) = \frac{freq(cluster_j)}{\sum_j freq(cluster_j)} \tag{3}$$

GBE has the property of being low when many clusters are condensed in a network. In a financial market, when the market is stable, the prices of each asset move freely to some extent, while when a kind of major event occurs and the market becomes unstable, the prices of many assets will fall or rise uniformly. Therefore, GBE can express stability in the financial market, and if the value of GBE falls sharply, it can be said that the market has moved from a stable state to an unstable state.

### 2.2 Differential networks

To construct differential networks, we first create differential matrixes from the distance matrixes. The distance matrixes are created in the same way as described in the procedure for constructing the co-occurrence networks, and then the difference matrixes are created by taking the difference at each time step.

Let the distance matrix at time t be X_t and the difference matrix D_t is defined as follows.

$$D_t = X_t - X_{t-1}\ (t \geq 2) \tag{4}$$

The absolute value of the difference matrix forms a differential network by putting an edge on a pair of assets whose absolute value exceeds a certain threshold.

Differential network analysis is a method originally used in the field of bioinformatics[13], and in previous studies, it has been applied to gene regulatory networks in normal and cancer states to identify factors that contributed significantly to network changes as cancer progresses [15]. On a differential network, since edges are drawn over pairs whose relations have changed significantly, it is thought that nodes that have changed significantly will collect a large number of edges, which will then appear on the network as hubs. Although differential networks are used in bioinformatics, there are not so many examples of their use in other fields, including the field of finance.

In this study, we use differential networks to identify and visualize assets that have made significant changes in the whole financial market. Besides, when the number of hubs represented on the differential network is high, it can be considered that large-scale changes are taking place in financial markets.

## 3 Experimental Details

We construct co-occurrence and differential networks and calculate GBE based on the time-series data of various types of financial assets. We used 49 time-series data consisted of US stocks (S&P500), Japanese stocks (TOPIX), US treasury, Japanese government bonds, and exchange rates (dollar/yen) for the period from 01/01/2007 to 31/12/2007, with daily price. These financial data were obtained from Bloomberg.

The window width of the DTW distance calculation was set to 20 days (days that stock markets are open in a month), the threshold of the distance to draw the edge in the co-occurrence network was set to be less than 2.0, and the threshold of the difference of the distance to draw the edge in the differential network was set to be more than 1.0 in absolute value, and the distance away from the edge with a value greater than 1.0 was connected with the red edge and the distance closer with a value less than -1.0 was connected with the blue edge. For each node in the differential network, those with a degree greater than or equal to 3 are considered as hubs.

## 4 Results and Discussion

In the differential networks, the number of hubs consisting of blue edges (hereinafter referred to as "closer hubs") has increased rapidly since the end of February (Fig. 2), and the visualized figures (Fig. 4) show that hubs have been formed around Japanese stocks (green nodes) and US stocks (red nodes). The value of GBE has also fallen sharply at the same time. In the corresponding co-occurrence network, the Japanese and U.S. stock clusters were separated on February 27, 2007 (Fig. 2), and gradually joined together to form a densely packed cluster two weeks later (Fig. 3).

February 27, 2007 was known as the day of "the Chinese Stock Bubble of 2007". The fall was caused by rumors that the Chinese governmental economic authorities were to introduce varying policies that would restrict foreign investment. As a result, it is said to have caused declines and major unrest in almost every financial market in the world [17]. The combination of Japanese and U.S. stocks on the co-occurrence network and the increase in closer hubs centered on Japanese and U.S. stocks on the differential network can be considered consistent with the phenomenon of simultaneous global stock price declines in the real world. Here, the U.S. and Japanese markets are very far apart geographically, with a 13-hour time difference. Due to the time difference between the two markets, there is a discrepancy in the timing at which these common factors affect price movements. The method seems to be able to handle these time gaps well because it detects changes in a graph-based manner using the DTW distance, which is flexible in terms of time gaps. This method can be said to capture the relational changes in financial markets in an interpretable way.

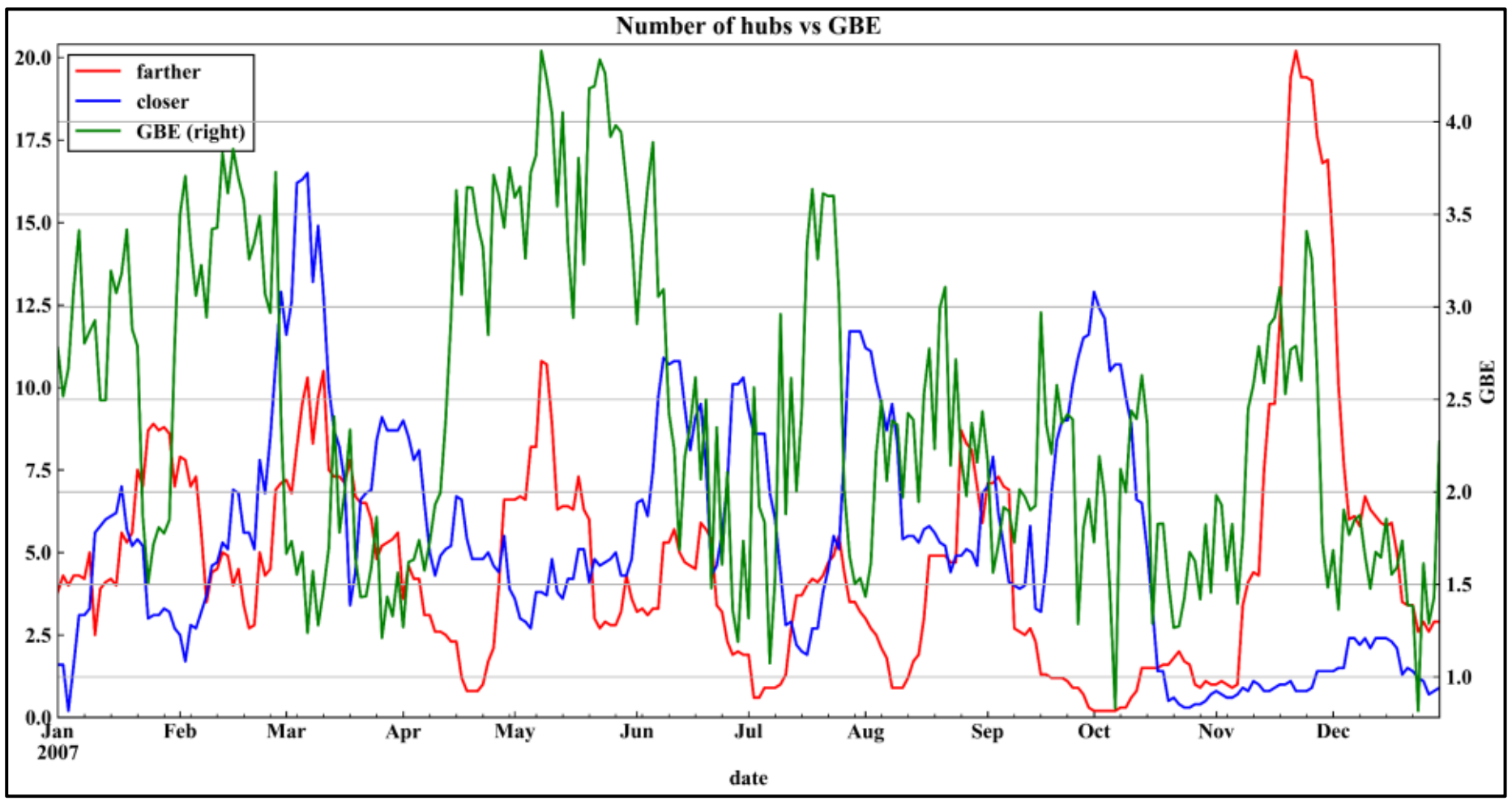

**Fig. 2.** The number of hub nodes. The red line means the number of father hubs, the blue line means the number of closer hubs, and the green line (right axis) means GBE.

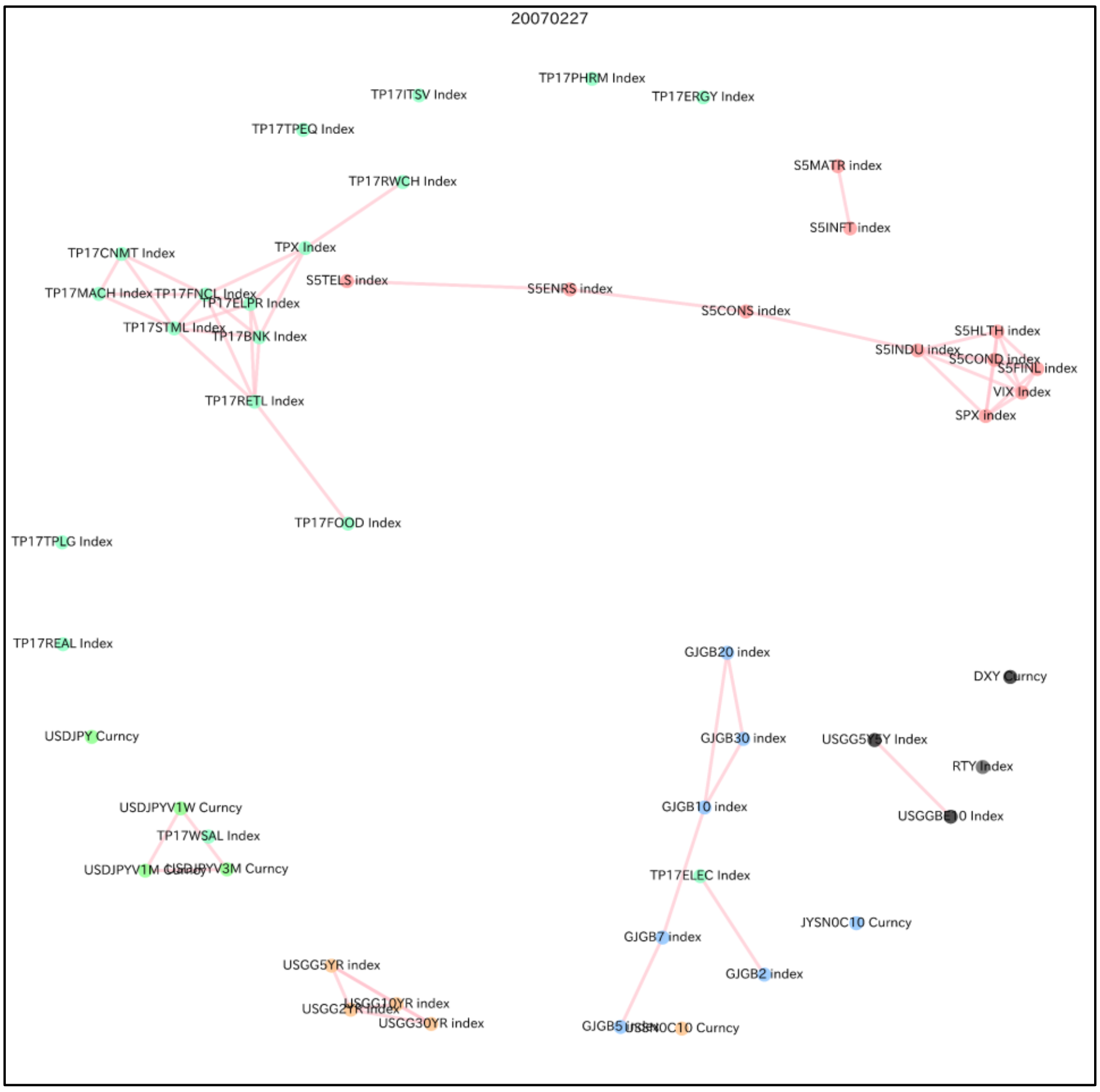

**Fig. 3.** Co-occurrence Network (Whole Image) on February 27, 2007. The colors of the nodes are as follows: Red: US stocks, Green: Japanese stocks, Orange: US Treasuries, Blue: Japanese government bonds, Green-yellow: dollar-yen exchange, Black: others.

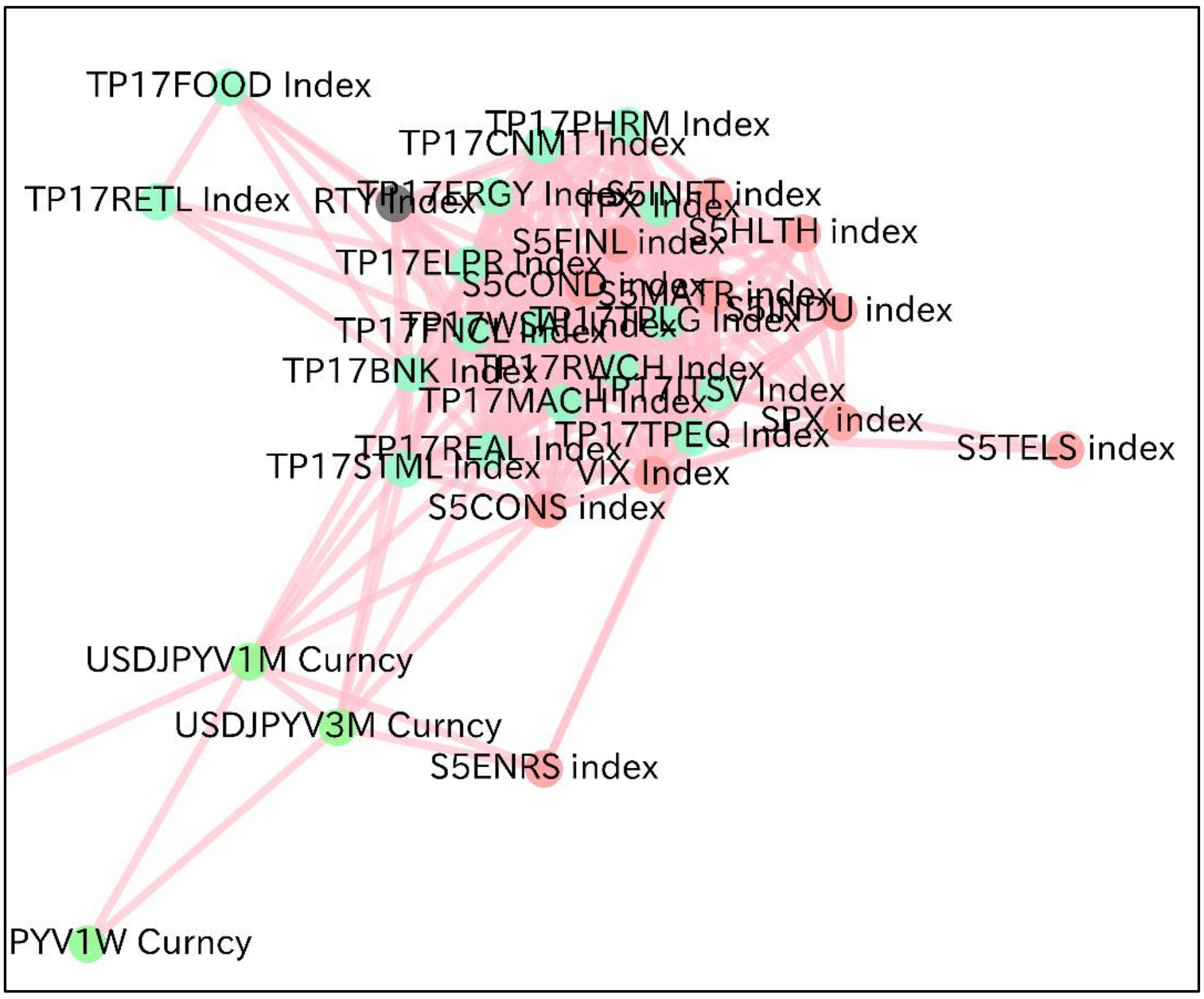

**Fig. 4.** Co-occurrence Network focusing on Japanese and US stocks cluster on March 13, 2007. The colors of the nodes are as follows: Red: US stocks, Green: Japanese stocks, Orange: US Treasuries, Blue: Japanese government bonds, Green-yellow: dollar-yen exchange, Black: others.

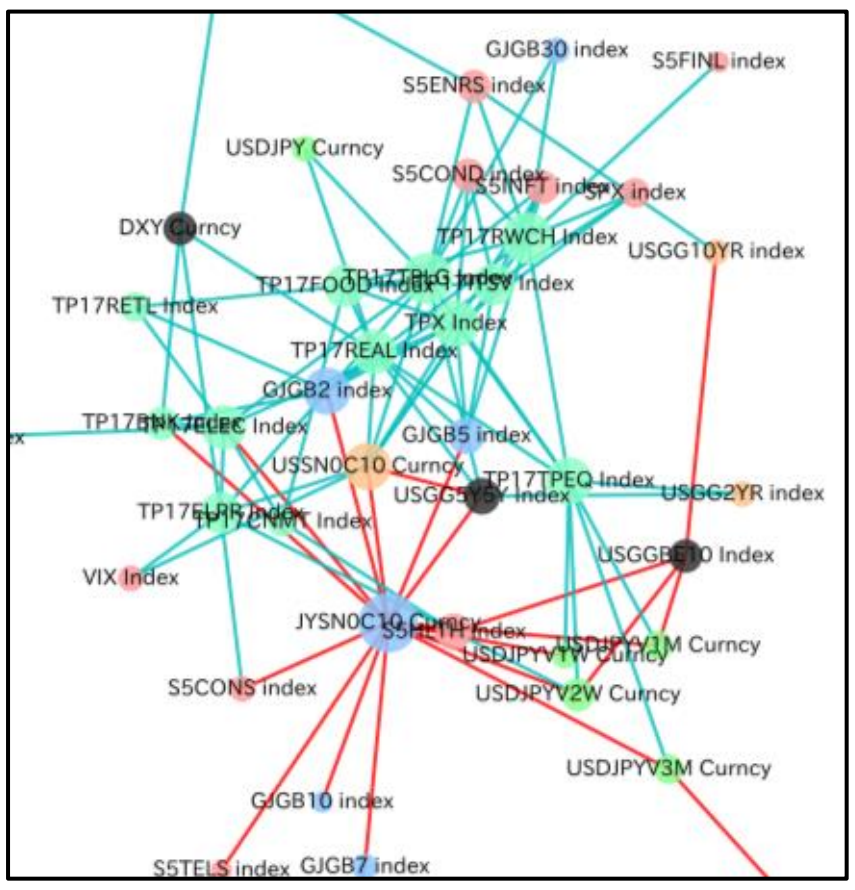

(a)

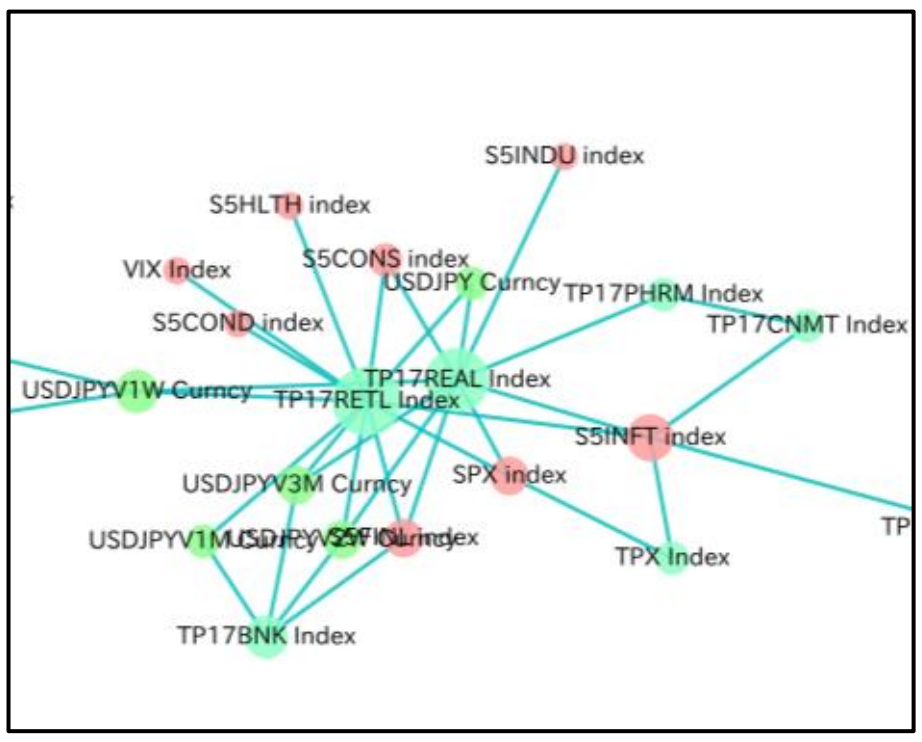

(b)

**Fig. 5.** (a)Differential Network focusing on Hubs on February 27, 2007. (b) Differential Network focusing on Hubs on March 13, 2007. The network consisted of blue edges is Closer Network, The network consisted of blue edges is Further Network. The colors of the nodes are as follows: Red: US stocks, Green: Japanese stocks, Orange: US Treasuries, Blue: Japanese government bonds, Green-yellow: dollar-yen exchange, Black: others.

Besides, the number of hubs consisting of red edges (hereafter referred to as further hubs), which signify that the distance has gone away, has also increased sharply through November, with GBE values rising. The breakdown of the hubs on the differential network is dominated by U.S. Treasuries Fig. 5(d), and in the co-occurrence network, a single cluster of tightly coupled U.S. Treasuries and JGBs was formed in early November(Fig. 5(a)), but the cluster gradually became loosely coupled(Fig. 5(b)) and separated into two clusters in mid to late November (Fig. 5(c)). During this period, the interest rates on short-term bonds were quite low because of the subprime mortgage crisis, which shifted funds from long-term bonds such as stocks and U.S. 10-year bonds to short-term bonds such as 2-year bonds. As a result, it is thought that the price movement of Japanese government bonds has come to be different from that of other Japanese government bonds.

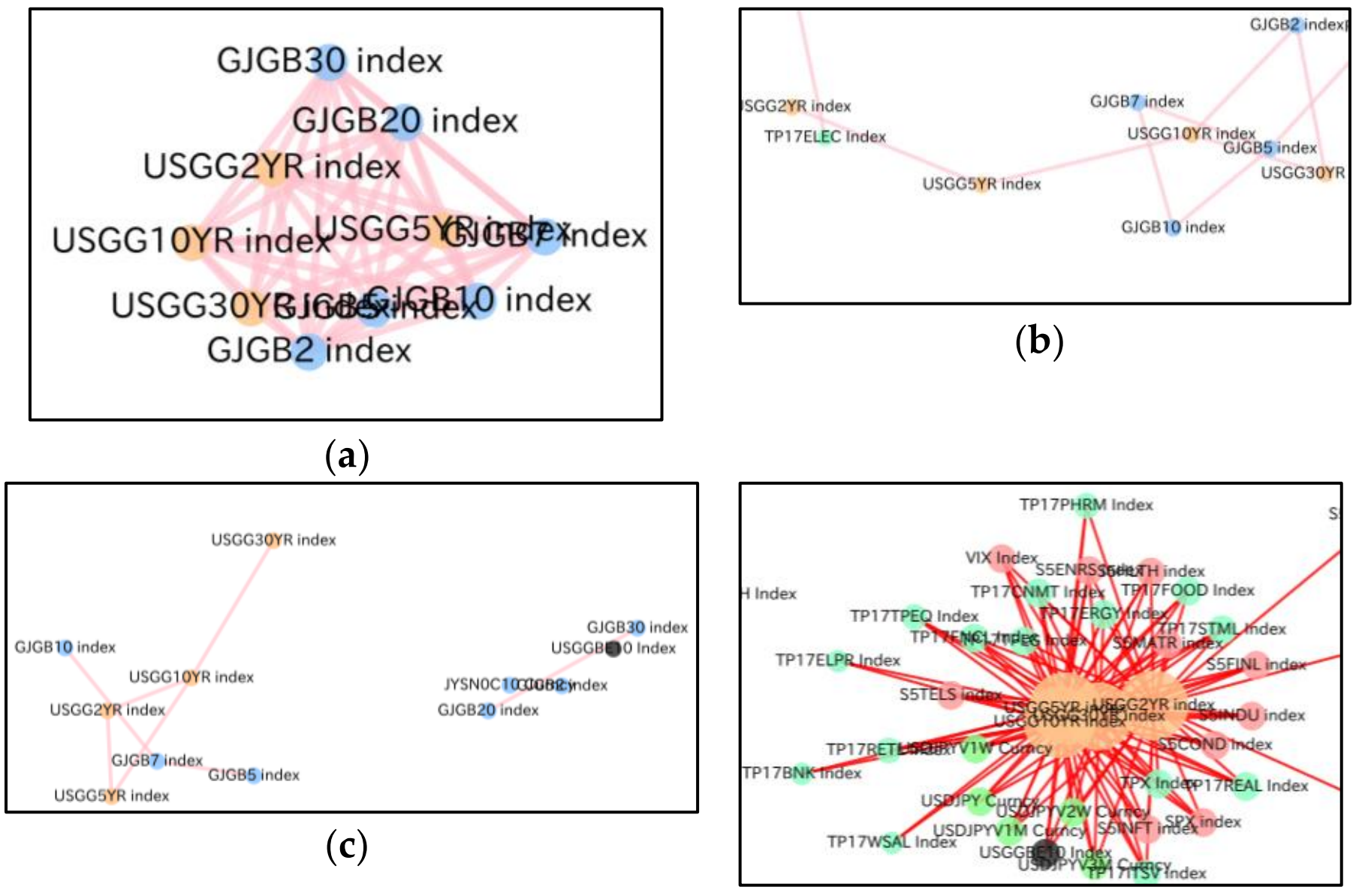

**Fig. 6.** (a)Co-occurrence Network focusing on Japanese and US bonds clusters on November 1, 2007. (b) Co-occurrence Network focusing on Japanese and US bonds clusters on November 13, 2007. (c)Co-occurrence Network focusing on Japanese and US bonds clusters on November 19, 2007. (d) Differential Network focusing on Japanese and US bonds clusters on November 13, 2007. The colors of the nodes are as follows: Red: US stocks, Green: Japanese stocks, Orange: US Treasuries, Blue: Japanese government bonds, Green-yellow: dollar-yen exchange, Black: others.

The GBE values and the number of hubs allow us to know when the structure of the network has changed significantly and its scale, while the differential network visualizes which assets are central to the changes in the network. Furthermore, the differential network allows us to focus on the assets that are central to the change and how they change on the network, as captured by the differential networks. The two aforementioned examples suggest that such an approach can assist in interpreting what changes are occurring in a multi-asset market. This set of procedures can be used as one of the criteria to assist fund managers in making operational decisions, such as reconfiguring portfolios of assets. For example, if we detect timing that is different from the market price movements normally expected from the level of GBE, we can consider why the change has occurred and if it is something that is long term, we can decide to cash out the asset and exit the market. Alternatively, if we can consider from the graph that the change in the relationship is short term, we could make the decision to take a mean reversion strategy. By buying and selling assets when the relationship between the two assets, which is usually expected, breaks down, we can buy and sell assets with the expectation that the relationship will eventually return in the long run, thus allowing us to aim for profit opportunities while limiting risk.

While the study shows high interpretability and can show qualitatively valid results because they are consistent with real-world events, the methodology does not aim to optimize operational decisions and it is up to the fund manager to decide how to link them to operational decisions. Therefore, there is no quantitative evaluation of how much profit this method generates, and it remains only a qualitative evaluation. Also, this study aims to support human decision-making and focuses on "explanation" rather than "detection" or "prediction" of time-series changes. Because it is a study on a new goal that differs from many existing studies, there are no existing methods to compare currently. Therefore, quantitative evaluation measures are required in the future.

Future research could include proposing uses and methods that lead to greater operational optimization and even quantitative evaluation while maintaining interpretability. Also, we use several hyperparameters, such as distance thresholds for constructing a network and order thresholds for being considered a hub in a differential network. These thresholds were set in a deterministic manner, but a methodology is required to optimize the thresholds to lead to more useful analysis.

In the current experiment, we focused on stocks, bonds, and foreign exchange, but we may be able to uncover new, previously unnoticed relationships by using it on more types of assets and indices, as well as on so-called alternative data, which is not traditionally used, in the future. By doing so, we may be able to discover new revenue opportunities. Also, the fact that clustering is performed on the network may allow us to apply a mean reversion strategy from a many-to-many relationship rather than a one-to-one relationship. Further investigation of the specific applications of this method in future research would have the potential to be more profitable.

## 5 Conclusion

This study detects changes in relationships in multi-asset by using indicators of GBE and the number of hubs in a differential network and visualizes them in the form of highly interpretable networks to help fund managers capture the changes taking place in real markets.

The Experimental results suggest that the co-occurrence network, GBE, and the differential network, can be applied to detect changes across multiple assets in financial markets in a sufficiently interpretable manner and that this is in line with the events of what is happening in the real world. GBE and differential networks' hub count metrics provide a bird's eye view of the changing multi-asset relationships. It is hoped that this will allow it to be used as one of the indicators in the management of multi-asset funds, of which there are not many types yet, to assist fund managers in their decision-making. So far, co-occurrence networks and GBE have been used only in studies of changes in consumer behavior in supermarkets, and differential networks have only been used in studies in the context of bioinformatics, and there have been few studies that have applied them to other fields, including finance. This study shows that it can be applied to financial markets as well.

### References

1. AHMAD, Subutai, et al. Unsupervised real-time anomaly detection for streaming data. *Neurocomputing*, 2017, 262: 134-147.J.P.
2. AMINIKHANGHAHI, Samaneh; COOK, Diane J. A survey of methods for time series change point detection. *Knowledge and information systems*, 2017, 51.2: 339-367.
3. Morgan Asset Management. The A.B.C. of multi-asset income investing. Available online: https://am.jpmorgan.com/hk/en/asset-management/per/insights/investment-ideas/multi-asset-income/ (Accessed on 28, 8, 2020)
4. SCHWENDNER, Peter, et al. Multi-asset correlation dynamics: Impact for specific investment strategies and portfolio risk. *In: TSAM Europe: Performance Measurement & Investment Risk,* London, United Kingdom, 1 March 2014. The Summit for Asset Management (TSAM), 2014.
5. MARKOWITZ, Harry. Portfolio selection. *Journal of Finance*, 1952, 7(1), 77-91.
6. PESKIN, K. Stuart. Evaluating Multi-Asset Strategies. *The Journal of Portfolio Management*, 2017, 44.2: 40-49.
7. AKOGLU, Leman; TONG, Hanghang; KOUTRA, Danai. Graph based anomaly detection and description: a survey. *Data mining and knowledge discovery,* 2015, 29.3: 626-688.
8. ONNELA, Jukka-Pekka, et al. Financial market-a network perspective. *In: Practical Fruits of Econophysics*. Springer, Tokyo, 2006. p. 302-306.
9. KUMAR, Sunil; NIVEDITA Deo. Correlation and network analysis of global financial indices. *Physical Review E,* 2012 86(2), 026101.

10. GIUDICI, Paolo; ALESSANDRO Spelta. Graphical network models for international financial flows. *Journal of Business & Economic Statistics.* 2016;34.1:128-38.
11. OHSAWA, Yukio. Graph-based entropy for detecting explanatory signs of changes in market. The Review of Socionetwork Strategies, 2018, 12.2: 183-203.
12. NARAOKA, Makoto; HAYASHI, Teruaki.; OHSAWA, Yukio. Constructing the time segment analysis method for planning strategies in supermarket. [Translated from Japanese.] *The Institute of Electronics, Information and Communication Engineers SIG-AI: Data Market V*, 2019, Vol.118, No.453, pp.61-65.
13. IDEKER, Trey; KROGAN, Nevan J. Differential network biology. *Molecular systems biology*, 2012, 8.1: 565.
14. HA, Min Jin; BALADANDAYUTHAPANI, Veerabhadran; DO, Kim-Anh. DINGO: differential network analysis in genomics. *Bioinformatics*, 2015, 31.21: 3413-3420.
15. XU, Ting, et al. Time-Varying Differential Network Analysis for Revealing Network Rewiring over Cancer Progression. *IEEE/ACM Transactions on Computational Biology and Bioinformatics*, 2019.
16. SAKOE, Hiroaki; CHIBA, Seibi. Dynamic programming algorithm optimization for spoken word recognition. *IEEE transactions on acoustics, speech, and signal processing*, 1978, 26.1: 43-49.
17. That's "This Day in History: The 2007 China Stock Market Crash." Available online: https://www.thatsmags.com/china/post/12485/this-day-in-history-the-2007-china-stock-market-crash (Accessed on 28, 8, 2020)